\documentclass[9pt,conference]{IEEEtran}
\IEEEoverridecommandlockouts

\usepackage{amsmath}
\usepackage{bm}
\usepackage{amssymb}
\usepackage{booktabs}
\usepackage{amsmath,amssymb,amsfonts}
\usepackage{algorithmic}
\usepackage{graphicx}
\usepackage{textcomp}
\usepackage{xcolor}
\usepackage{array}
\usepackage{booktabs}
\usepackage{subcaption}

\newcommand{\vect}[1]{\boldsymbol{#1}} % vector & matrix bold

\usepackage{soul} % strike \st
\usepackage{colortbl}
 \usepackage{multirow}
\usepackage{wrapfig}

\usepackage{cite}
\usepackage{amsmath,amssymb,amsfonts}
\usepackage{algorithmic}
\usepackage{graphicx}
\usepackage{textcomp}
\usepackage{xcolor}
\def\BibTeX{{\rm B\kern-.05em{\sc i\kern-.025em b}\kern-.08em
    T\kern-.1667em\lower.7ex\hbox{E}\kern-.125emX}}
\begin{document}

\title{Multiview Canonical Correlation Analysis for \\ Automatic Pathological Speech Detection\\
\thanks{This work was supported by the Swiss National Science
Foundation project CRSII5\_202228 on “Characterisation of
motor speech disorders and processes".}
\thanks{\textsuperscript{*}Equal contribution}
}

\author{\IEEEauthorblockN{Yacouba Kaloga\textsuperscript{*}, Shakeel A. Sheikh\textsuperscript{*}, Ina Kodrasi}
\IEEEauthorblockA{
% \textit{Signal Processing} \\
\textit{Idiap Research Institute, Martigny, Switzerland} \\
\{yacouba.kaloga, shakeel.sheikh, ina.kodrasi\}@idiap.ch}
}

% \author{\IEEEauthorblockN{Yacouba Kaloga }
% \IEEEauthorblockA{
% % \textit{Signal Processing} \\
% \textit{Idiap Research Institute}\\
% Martigny, Swtzerland \\
% yacouba.kaloga@idiap.ch
% }
% \and
% % Signal Processing for Communication Group, Idiap Research Institute, Martigny, Switzerland
% \IEEEauthorblockN{Shakeel A. Sheikh}
% \IEEEauthorblockA{
% % \textit{Signal Processing} \\
% \textit{Idiap Research Institute}\\
% Martigny, Swtzerland  \\
% shakeel.sheikh@idiap.ch
% }
% \and
% \IEEEauthorblockN{Ina Kodrasi}
% \IEEEauthorblockA{
% % \textit{Signal Processing} \\
% \textit{Idiap Research Institute}\\
% Martigny, Swtzerland \\
% ina.kodrasi@idiap.ch
% }}
% \and
% % \IEEEauthorblockN{5\textsuperscript{th} Given Name Surname}
% % \IEEEauthorblockA{\textit{dept. name of organization (of Aff.)} \\
% % \textit{name of organization (of Aff.)}\\
% % City, Country \\
% % email address or ORCID}
% % \and
% % \IEEEauthorblockN{6\textsuperscript{th} Given Name Surname}
% % \IEEEauthorblockA{\textit{dept. name of organization (of Aff.)} \\
% % \textit{name of organization (of Aff.)}\\
% % City, Country \\
% % email address or ORCID}
% }

\maketitle

\begin{abstract}
   Recently proposed automatic pathological speech detection approaches rely on spectrogram input representations or wav2vec2 embeddings. 
   These representations may contain pathology-irrelevant uncorrelated information, such as changing phonetic content or variations in speaking style across time, which can adversely affect classification performance.
   To address this issue, we propose to use Multiview Canonical Correlation Analysis (MCCA) on these input representations prior to automatic pathological speech detection. 
    Our results demonstrate that unlike other dimensionality reduction techniques, the use of MCCA leads to a considerable improvement in pathological speech detection performance by eliminating uncorrelated information present in the input representations. 
    Employing MCCA with traditional classifiers yields a comparable or higher performance than using sophisticated architectures, while preserving the representation structure and providing interpretability.
\end{abstract}


\section{Introduction}

Speech production is a complex process where the brain executes a series of sequences involving various sensorimotor, muscle, and articulator processes.
The execution of these sequences can be affected by neurodegenerative disorders such as Parkinson's disease (PD)~\cite{damico2010handbook}, which can result in pathological speech characterized by imprecise articulation, insufficient prosody, and other abnormal speech patterns~\cite{tjaden08, yunusova2008articulatory}. 
In clinical settings, diagnosis of neurological impairments is typically done through auditory-perceptual tests alongside other meta data such as genetic information and visual cues~\cite{updrs}. 
However, diagnostic accuracy can vary among clinicians depending on their experience, implicit biases, as well as their condition during the diagnosis~\cite{pernon2022perceptual}. To address this issue, researchers are exploring various automatic pathological speech detection  approaches. 
These approaches mainly differ in the input representations and the classifers that are exploited.

In traditional automatic pathological speech detection approaches based on machine learning, handcrafted acoustic features often inspired by clinical knowledge are fed to classical algorithms, such as e.g., support vector machines (SVMs)~\cite{kodrasi2020spectro}, multi layer perceptrons (MLPs)~\cite{Farhadipour18}, or random forests~\cite{Joshy22}. 
Various feature sets have been examined, including openSMILE~\cite{narendra20}, Mel-frequency cepstral coefficients~\cite{janbakhshi_ua}, or sparsity-based features~\cite{kodrasi2020spectro}.
Despite the reported promising performance, hand-crafted features have a limited capability to comprehensively capture nuances, cues, and complexities of pathological speech. 

Deep learning (DL)-based approaches on the other hand have the potential to capture more abstract and subtle cues of pathological speech. 
These approaches commonly employ spectrogram input representations such as the short-time Fourier transform~(STFT) or Mel spectrograms~\cite{janbakhshi_stft, kodrasi2020spectro}. 
Spectrograms capture both temporal and spectral information, making them well-suited for analyzing and interpreting speech signals~\cite{slp_book}.
Approaches exploiting spectrogram input representations often rely on complex architectures~\cite{Joshy22}. For example, convolutional neural networks~(CNNs) have been exploited in~\cite{vasquez2017convolutional, zaidi2021cnn} to leverage the two-dimensional structure of the time-frequency representations. 
In~\cite{kumar23}, long short-term memory (LSTMs) networks are used for their capability to capture long-range dependencies. 
Despite the higher performance reported by these approaches compared to approaches using handcrafted features, they no longer represent the state-of-the-art in the field. 
Today, approaches relying on self-supervised learning (SSL) models such as wav2vec2 (w2v2) are preferred~\cite{baevski2020wav2vec, w2v2_pd, sheikh2024impact}. 
These models leverage a vast collection of available audio data to learn embeddings which enable unprecedented performance for several downstream tasks~\cite{superb}. Despite their performance, interpreting SSL features remains challenging, discouraging their deployment in clinical practice. 
Consequently, approaches based on spectrogram input representations remain relevant and efforts are still directed at improving their performance.

Speech representations such as spectrograms and w2v2 embeddings include cues about speaker identity, phonetic content, speaking style, and emotional state. However, these cues can be irrelevant or even detrimental for pathological speech detection. To mitigate this issue, supervised adversarial and non-adversarial training methods have been proposed to suppress speaker identity cues\cite{janbakhshi2021supervised,janbakhshi2022adversarial}, while dimensionality reduction techniques such as principal component analyses (PCA) and linear discriminant analysis have been used to remove other redundant information \cite{kacha20, ARJMANDI20123}. Although dimensionality reduction techniques are well-studied for handcrafted acoustic features, they remain under-explored for DL-based input representations.
This is because DL-based approaches are expected to be powerful enough to ignore pathology-unrelated cues in input representations. However, due to the limited data typically available for this task, effectively learning to do so remains challenging.

In this paper, we propose to use multiview canonical correlation analysis (MCCA) \cite{kettering} to remove redundant information from input speech representations prior to training pathological speech detection approaches. 
While MCCA has been extensively investigated in the context of computer vision~\cite{deep_cca},
% [https://arxiv.org/pdf/1811.12345];
recommendation systems~\cite{chen_graph_2019}, silent speech recognition~\cite{kim17g_interspeech}  and various classification tasks~\cite{kaloga_variational_2021}, to the best of our knowledge, it has never been incorporated in pathological speech detection.
MCCA, a multiview version of PCA, finds a common representation across multiple views, which is often more effective for classification or clustering than using any single view or their concatenation \cite{Chen19}.
We hypothesize that unlike pathology-discriminant cues, pathology-irrelevant cues such as changing phonetic content or speaking style are uncorrelated over time. 
By considering different chunks of an utterance as separate views of the speaker within the MCCA framework, MCCA representations preserve the correlated pathology-discriminant cues while suppressing the uncorrelated pathology-irrelevant cues. 

 This approach maintains the structure of the representation, enabling state-of-the-art performance with simpler models, like MLPs, without losing interpretability.

\section{Multiview Canonical Correlation Analysis}
\label{sec:mcca}
%For completeness, this section provides a brief overview of the formulation and derivation of MCCA.

Elements of a dataset can be described from various "viewpoints", such as images of the same  object taken from different angles or the perception of an event acquired with two or more sensory inputs.
The representations of the same element from different perspectives can be considered as 
distinct views.
The objective of MCCA is to find 
a shared low-dimensional representation from distinct views of the element.

Given two views $X_1 \in \mathbb{R}^{f \times t_1}$ and $X_2 \in \mathbb{R}^{f \times t_2}$, we aim to find the optimal projectors $U_1 \in \mathbb{R}^{t_1 \times t}$ and $U_2\in \mathbb{R}^{t_2 \times t}$, where $t \le \min{(t_1, t_2)}$, such that the correlation between $X_1 U_1$ and $X_2 U_2$ is maximized.
This objective can be expressed as the optimization problem 

\begin{equation}
\label{cca_objective}
\min_{U_1,U_2} |\!|X_1 U_1 - X_2 U_2 |\!|_2^2 
\text{ s.t. } U_m^{T}(X_m^{T} X_m)U_m = I_{t},
\end{equation}
with $m \in \{1,2\}$ and $I_{t}$ being the $t \times t$--dimensional identity matrix.
The solution to~(\ref{cca_objective}) is obtained through an eigendecomposition, as described in \cite{Chen19}. Extending this approach to multiview data presents challenges, as maximizing the pairwise correlation between $M>2$ different views is NP-hard~\cite{Rup13}. 
The MAXVAR relaxation of this problem~\cite{kettering} seeks a shared low-dimensional representation $S \in \mathbb{R}^{f \times t}$ that closely approximates each low-dimensional projection $X_mU_m$, where $m = 1, \dots, M$. 
This leads to MCCA, which is formulated as

\begin{equation}
\label{mcca_objective}
 %   \begin{split}
    \min_{S,(U_m)_{m=1 \dots M}}  \sum_{ \substack{m=1}  }^M |\! |X_m U_m  - S  |\!|_2^2
     \text{ s.t } S^{T}S =  {I}_{t}.
\end{equation}

The optimal representation ${S^{*}}$ solving~(\ref{mcca_objective}) is obtained by extracting the columns corresponding to the $m$-leading eigenvectors of the matrix $\sum_{m=1}^{M} X_m^{T}(X_m X_m^{T})^{-1}X_m$ as described in~\cite{Chen19}\footnote{Note that $X_m X_m^T$, the covariance matrix of view $m$, is typically regularized as $X_m X_m^T +10^{-4} I_{t_m}$ for numerical stability.}, whereas the optimal projection matrices ${U_m^{*}}$ are computed as ${U_m^{*}} = (X_m X_m^{T})^{-1}X_mS^{*^T}$.
As shown in~\cite{kaloga_variational_2021, Chen19}, using the low-dimensional representation $S^{*}$ instead of the individual views $X_m$ (or their concatenation) yields an advantageous performance.

\section{Multiview Canonical Correlation Analysis \\ for Pathological Speech Detection}

We consider a neurotypical and pathological speech dataset containing $n$ elements, denoted as $\{\vect{x}_{i}\}_{i=1}^n$. Each element $\textbf{x}_i \in \mathbb{R}^{F \times T_i}$ represents a segment of speech data as e.g., a spectrogram or w2v2 embedding, where $F$ indicates the feature dimensionality and $T_i$ the number of time frames. Each representation is labeled as neurotypical or pathological.
The number of time frames $T_i$ can be different among the different elements of the dataset.
However, in order to increase the number of elements in the dataset, we consider representations of fixed-size segments of speech extracted from full utterances as in~\cite{janbakhshi2021supervised, janbakhshi_ua}.
State-of-the-art literature directly uses these representations to train pathological speech detection models. 
However, these representations contain extensive pathology-irrelevant cues
related to the varying phonetic content, speaking style, or emotional
state of the speaker, which can be detrimental to pathological speech detection performance. Our assumption in this paper is that some
irrelevant cues for pathological speech detection do not exhibit
temporal self-correlation. Hence, to remove uncorrelated irrelevant
cues from these representations, we propose to incorporate MCCA as described in the following.

To use MCCA for pathological speech detection, we divide the $F \times T_i$--dimensional representation $\vect{x}_i$ into $M$ chunks $\vect{x}_i^{(m)} \in \mathbb{R}^{ F \times \lfloor \frac{T_i}{M} \rfloor}$, with $M$ being a user-defined parameter~(cf. Section~\ref{sec:chunks}), $m \in {1,\dots,M}$, and $\lfloor \cdot \rfloor$ denoting the floor operator. 
% \begin{equation}
%     \vect{x}_i^{(m)} \in \mathbb{R}^{ F \times \lfloor \frac{T_i}{M} \rfloor}
% \end{equation}
If $M$ does not divide $T_i$, the remaining time frames are discarded. 
We consider these chunks to be the $M$ distinct views (denoted by $X_m$ in Section~\ref{sec:mcca}) in the context of MCCA.
% representation $\vect{x}_i^{(M)} $ will belong to $\mathbb{R}^{F \times T_i[M]}$ where $T_i[M]$ stand for the rest of euclidian divison of $T_i$ by $M$}
\begin{figure}[t!]
% \vspace*{-0.5cm}  
\hspace*{-0.4cm}  
    \centering \includegraphics[scale=0.4]{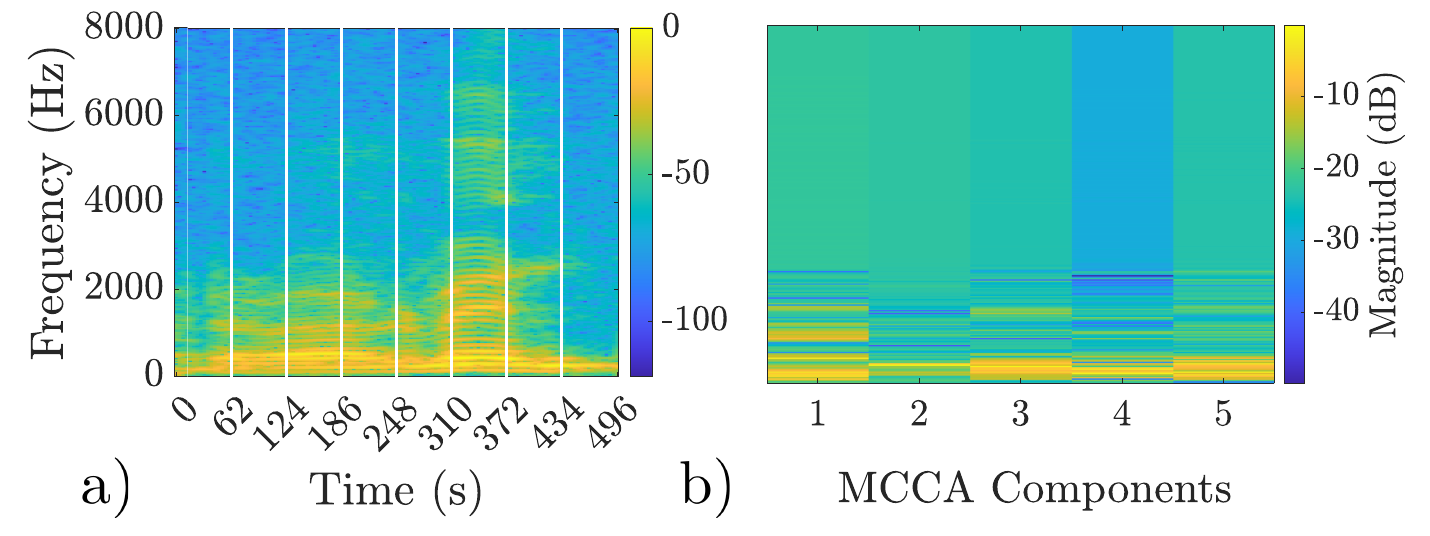}
     %\vspace{-0.3cm}
    \caption{(a) Spectrogram of a $500$~ms long speech segment with $257$ frequency bins representing the frequency range from $0$~Hz to $8000$~Hz. The vertical lines denote the boundary of $M = 8$ chunks, with each chunk considered to be a single view of the speech segment. (b) Representation after applying MCCA, corresponding to the same $257$ bins and containing $T = 5$ components.} 
    \vspace{-0.5cm}
    \label{fig:mcca}
\end{figure}
After applying MCCA  (cf.~(\ref{mcca_objective})), the original $F \times T_i$--dimensional representation $\mathbf{x}_i$ is reduced to the $F \times T$--dimensional representation $\vect{x}_i^{*}$ (denoted by $S^*$ in Section~\ref{sec:mcca}), with $T$ representing the number of MCCA components.
These dimensionality-reduced representations of speech segments can then be used as input to classification models such as MLP or LGBM.
The choice of the number of chunks $M$ reflects the time span along which relevant cues are expected to be correlated while irrelevant cues are expected to be uncorrelated.
By carefully selecting $M$, we can eliminate irrelevant cues from input representations and improve the performance of pathological speech detection approaches.
However, different representations exhibit different characteristics and correlation spans, making it challenging to derive a general optimal number of chunks $M$ from a theoretical perspective. 
Experimental results in Section~\ref{sec:chunks} provide insights into this matter.

Fig.~\ref{fig:mcca}a depicts an exemplary spectrogram of a speech segment (i.e., the representation used by state-of-the-art approaches based on the STFT). For a user-defined number of chunks $M=8$, the vertical lines depicted in Fig.~\ref{fig:mcca}a show the boundaries between the different chunks that are considered to be the different views of this spectrogram.
Applying MCCA with $T=5$ components to these chunks yields the representation in Fig.~\ref{fig:mcca}b.
%Instead of using the original representation in Fig.~\ref{fig:mcca}a, we propose to use the MCCA output in Fig.~\ref{fig:mcca}b as input representations to pathological speech detection models.
As expected, MCCA preserves the feature structure  (i.e., along
the F dimension) and contains most of its variance in the low frequency components.
Since MCCA preserves the  feature structure, it enables the identification of features that influence the decision of classifiers. 
This is particularly important when using spectrogram representations, where the $F$ features represent the energy in different frequency bins.

\section{Experimental Settings}
\label{overview}
In the following, we present the various experimental settings used for generating the experimental results in Section~\ref{chunkssize}.

\emph{Dataset.} \enspace 
 We use Spanish recordings from a group of $50$ patients diagnosed with Parkinson’s disease ($25$ males, $25$ females) along with $50$ neurotypical speakers ($25$ males, $25$ females) from the PC-GITA database~\cite{orozco2014new}. 
 
Spontaneous speech recordings of speakers discussing their day are used. Recordings are downsampled to $16$~kHz and segmented into $500$~ms segments with a $50$\% overlap prior to computing input representations.
The average length of the total available speech material for each speaker is $47.1$ s.

\begin{figure}
\vspace{-0.4cm}
% \vspace*{-0.3cm} 
\hspace*{-0.8cm}  
    \centering
    \includegraphics[scale=0.35]
    {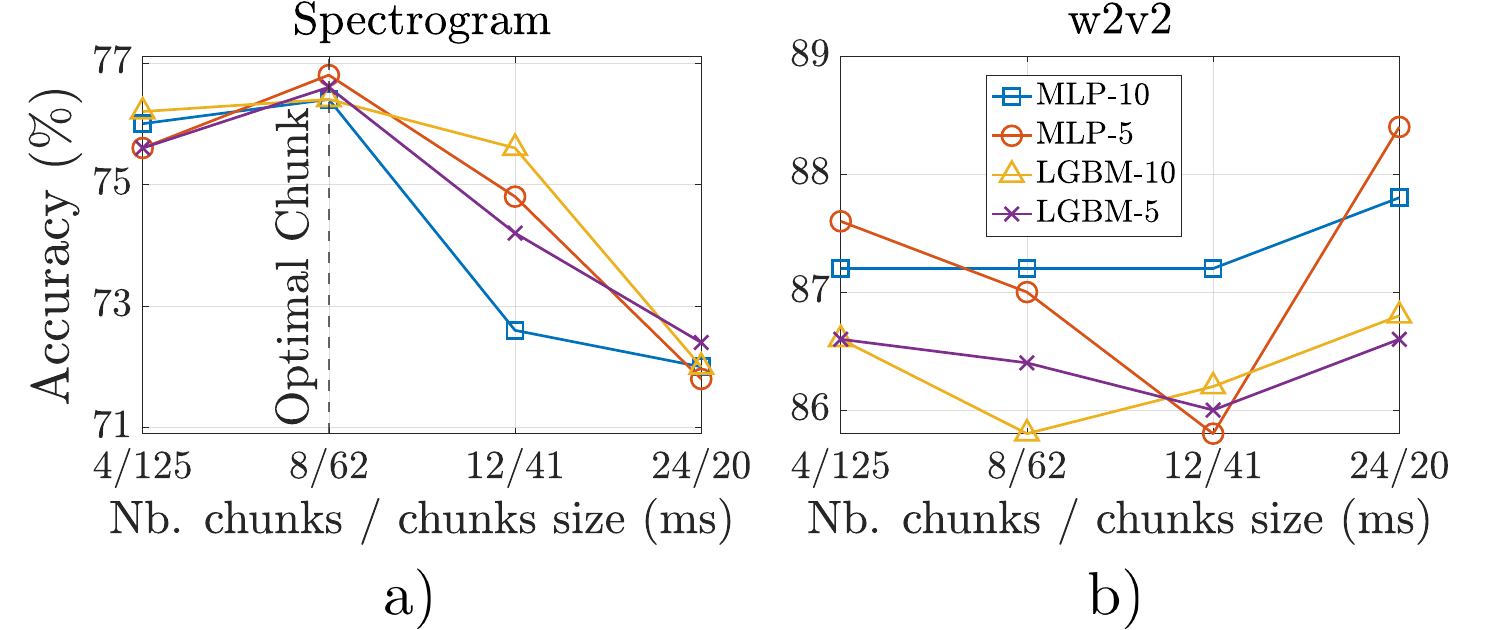}
    \caption{MLP and LGBM performance for different chunk sizes $M$ with $5$ and $10$ MCCA components using (a) spectrogram and (b) w2v2 embeddings.}
     \vspace{-0.5cm}
    \label{fig:enter-label}
\end{figure}

\emph{Input representations and detection models.} \enspace 
We investigate the applicability of the proposed MCCA approach to both spectrogram and w2v2 input representations.
For the spectrogram representations, the STFT is computed using a $32$~ms Hamming window with a hop size of $4$~ms, resulting in $257 \times 126$--dimensional representations.
For the w2v2 representations, we extract embeddings from the first layer of the transformer module of the XLRS53 version of w2v2 as in~\cite{sheikh2024impact}, resulting in
$1024 \times 24$--dimensional representations.
These input representations (and their MCCA versions) are used to train two different pathological speech detection models; an MLP, which is a common choice in pathological speech research~\cite{janbakhshi_ua}, and LGBM, which is known for its superior performance with careful parameter selection~\cite{lgbm}.

\emph{Evaluation.} \enspace
Evaluation is done in a speaker-independent stratified $10$-fold cross-validation framework. At each fold, $80\%$, $10\%$, and $10\%$
of the data is used for training, validation, and testing, respectively. 
To account for the effect of random initialization of the models when training, we train each model with $5$ different seeds and report the average and standard deviation of the
performance across these different seeds. 
The performance is evaluated in terms of speaker-level accuracy, which is computed
through soft voting of the probability of decisions for all segments belonging to each speaker.

\emph{Training.} \enspace 
For each input representation and detection model, we conduct hyperparameter tuning on the validation set (as specified below), with specific grid searches for each model focusing on key parameters to balance training time and performance.

\begin{itemize}

    \item MLP: Hyperparameters include the number of hidden layers $\{2,3,4\}$ with $\{64,128\}$ units per layer. The maximum number of iterations is fixed at $3000$. %This results in a total of 6 combinations for exploration.

    \item LGBM: Hyperparameters include $\textit{num\_leaves} =31$, $\textit{min\_child\_samples} \in \{2,3\}$, $\textit{max\_depth} \in  \{20,30\}$, $\textit{n\_estimators} \in \{400, 500, 600\}$, $\textit{colsample\_bytree} \in \{0.1, 0.2\}$, $\textit{learning\_rate} = 0.01$, $\textit{is\_unbalanced} = \textit{True}$, and $\textit{boosting type} = \textit{Dart}$. %This yields a total of 48 combinations for exploration. 

\end{itemize}
 All remaining MLP and LGBM hyperparameters are set to default values from their respective ~\textit{scikit-learn}~1.2.2 and~\textit{LightGBM}~4.2.0 libraries.

\section{Experimental Results and Discussion}

In the following, the pathological speech detection performance achieved when applying MCCA to input representations is extensively investigated.
In Section~\ref{sec:chunks} we analyze the impact of the number of chunks $M$ for computing the MCCA representations on the detection performance.
In Section~\ref{sec:ccapca} the detection performance achieved using MCCA is compared to the traditionally used PCA.
In Section~\ref{sec:impo} the performance using MCCA is further improved through feature selection and insights on interpretability are provided.

\label{chunkssize}

\begin{table}[ht!]
 % \vspace*{-0.2cm} 
  \caption{MLP and LGBM performance using PCA and MCCA with different number of components on spectrogram representations.}
  
  %\caption{Performance using spectrogram inputs with different number of PCA or CCA (with $M=8$ chunks) components.}
  \label{tab:res_spec}
  %\centering
    \hskip0.5cm
\scalebox{0.95}{\begin{tabular}{c|c|ccccc}
  
    \toprule
     \multicolumn{2}{c}{} 
     & 
    \multicolumn{5}{c}{\textbf{Components}}     \\
    \midrule
    % $1$                       & $/10$ & $-20$~~~             \\
     \textbf{Models} & \textbf{Features}&  1& 2&3 &5    &  10        \\
    \midrule
     \multirow{4}{*}{MLP}  & \multirow{2}{*}{PCA}  &  $64.40 $ &$67.20$&  $68.20$ &$69.20$ & $71.00$         \\
     & & \color{gray!95}$\pm 3.44$ &{\color{gray!95}$\pm 3.43$}&  {\color{gray!95}$\pm 1.17$} &{\color{gray!95}$\pm2.71$} & {\color{gray!95}$\pm1.79$}           \\
    \cmidrule(lr){2-7}   
      % & CCA-8& $-$ &$74.4\pm2.5$&  $76.2\pm3.4$ &$76.8\pm2.3$ & $76.4\pm3.0$           \\
       & \multirow{2}{*}{MCCA} & $68.20 $ &$74.40$&  $76.20$ &$76.8$ & $76.40$       \\
      & & \color{gray!95}$\pm 1.60$ &{\color{gray!95}$\pm2.50$}&  {\color{gray!95}$\pm3.43$} &{\color{gray!95}$\pm2.32$} & {\color{gray!95}$\pm3.00$}           \\
      
      \midrule
     \multirow{4}{*}{LGBM}   &\multirow{2}{*}{PCA} &  $68.60$ &$70.20$&  $70.20$ &$70.60$ & $70.60$           \\
      & &  {\color{gray!95}$\pm 1.02$} &{\color{gray!95}$\pm1.60$}&  {\color{gray!95}$\pm2.14$} &{\color{gray!95}$\pm2.50$} & {\color{gray!95}$\pm2.65$}       \\

    \cmidrule(lr){2-7}   
      
       % & CCA-8& $-$ &$75.8\pm1.3$&  $75.6\pm0.5$ &$76.6\pm1.6$ & $76.4\pm1.2$              \\
       & \multirow{2}{*}{MCCA}  & $72.20$ &$75.80$&  $75.60$ &$76.60$ & $76.40$            \\
       & & {\color{gray!95}$\pm1.94$} &{\color{gray!95}$\pm1.33$}&  {\color{gray!95}$\pm0.49$} &{\color{gray!95}$\pm1.62$} & {\color{gray!95}$\pm1.20$}           \\
    \bottomrule
  \end{tabular}}
  % \vspace{-0.3cm}
\end{table}
\begin{table}[ht!]
  \caption{MLP and LGBM performance using PCA and MCCA with different number of components on w2v2 representations.} 
 
  \label{tab:res_w2v2}
  %\centering
   \hskip 0.4cm
  \scalebox{0.97}{\begin{tabular}{c|c|ccccc}
  
    \toprule
     \multicolumn{2}{c}{}     &    \multicolumn{5}{c}{\textbf{Components}}   \\
    \midrule
    % $1$                       & $/10$ & $-20$~~~             \\
     \textbf{Models} & \textbf{Features}&  1& 2&3 &5    &  10         \\
    \midrule
     \multirow{4}{*}{MLP}  & \multirow{2}{*}{PCA}  &  $83.20 $ &$84.60$&  $84.60$ &$85.60$ & $87.20$        \\
     
     & & \color{gray!95}$\pm 1.94$ &{\color{gray!95}$\pm 1.62$}&  {\color{gray!95}$\pm 2.24$} &{\color{gray!95}$\pm2.24$} & {\color{gray!95}$\pm 1.17$}          \\

    \cmidrule(lr){2-7} 
     
      % & CCA-8& $-$ &$74.4\pm2.5$&  $76.2\pm3.4$ &$76.8\pm2.3$ & $76.4\pm3.0$           \\
       & \multirow{2}{*}{CCA} & $82.80 $ &$84.40$&  $86.60$ &$88.40$ & $87.80$     \\
       
      & & \color{gray!95}$\pm 2.04$ &{\color{gray!95}$\pm1.83$}&  {\color{gray!95}$\pm1.62$} &{\color{gray!95}$\pm1.55$} & {\color{gray!95}$\pm1.47$}       \\

      \midrule
     \multirow{4}{*}{LGBM}   &\multirow{2}{*}{PCA} &   $86.00$ & $85.40$& $85.20 $  & $85.80$  &       $85.60$  \\
     
      & &  {\color{gray!95}$\pm 1.41$} &{\color{gray!95}$\pm0.49$}&  {\color{gray!95}$\pm1.17$} &{\color{gray!95}$\pm1.17$} & {\color{gray!95}$\pm1.72$}         \\

    \cmidrule(lr){2-7} 
      
       % & CCA-8& $-$ &$75.8\pm1.3$&  $75.6\pm0.5$ &$76.6\pm1.6$ & $76.4\pm1.2$              \\
       & \multirow{2}{*}{CCA} & $86.60 $ &$86.40$&  $86.60$ &$86.60$ & $86.80$            \\
       
       & & {\color{gray!95}$\pm1.74$} &{\color{gray!95}$\pm1.67$}&  {\color{gray!95}$\pm0.49$} &{\color{gray!95}$\pm0.49$} & {\color{gray!95}$\pm1.72$}          \\

    \bottomrule
  \end{tabular}}
  % \vspace{-0.5cm}
\end{table}
\subsection{Impact of the number of chunks}
\label{sec:chunks}

As previously mentioned, to apply MCCA to input representations, we first need to determine the optimal number of chunks \(M\), which corresponds to the time scale over which correlation is maximized. 
To this end, we evaluate the performance of MLP and LGBM classifiers using spectrogram and w2v2 input representations for different chunk sizes, i.e., \(M=4\) ($125$ ms), \(M=8\) ($62$ ms), \(M=12\) ($41$ ms), and \(M=24\) ($20$ ms). 
To ensure that the derived optimal number of chunks for each representation are not specific to a single number of MCCA components, we also consider two different numbers of components, i.e., $5$ and $10$. The number of chunks, along with the other hyperparameters described in Section~\ref{overview}, are optimized on the validation set. 
Such a procedure results in \(M=8\) as the optimal $M$ for spectrogram representations and \(M=24\) as the optimal $M$ for w2v2 representations, regardless of the classifier used and number of MCCA components. 
Fig.~\ref{fig:enter-label} shows the performance on the test set for each \(M\) for all considered input representations, classification models, and number of MCCA components, where it is confirmed that \(M=8\) and \(M=24\) yield the best performance for spectrogram and w2v2 represetations, respectively.

\begin{figure}[ht]
\vspace*{-0.4cm} 
\hspace*{-0.36cm}
% \hspace*{cm}  
    \centering
    \includegraphics[scale=0.4]
    {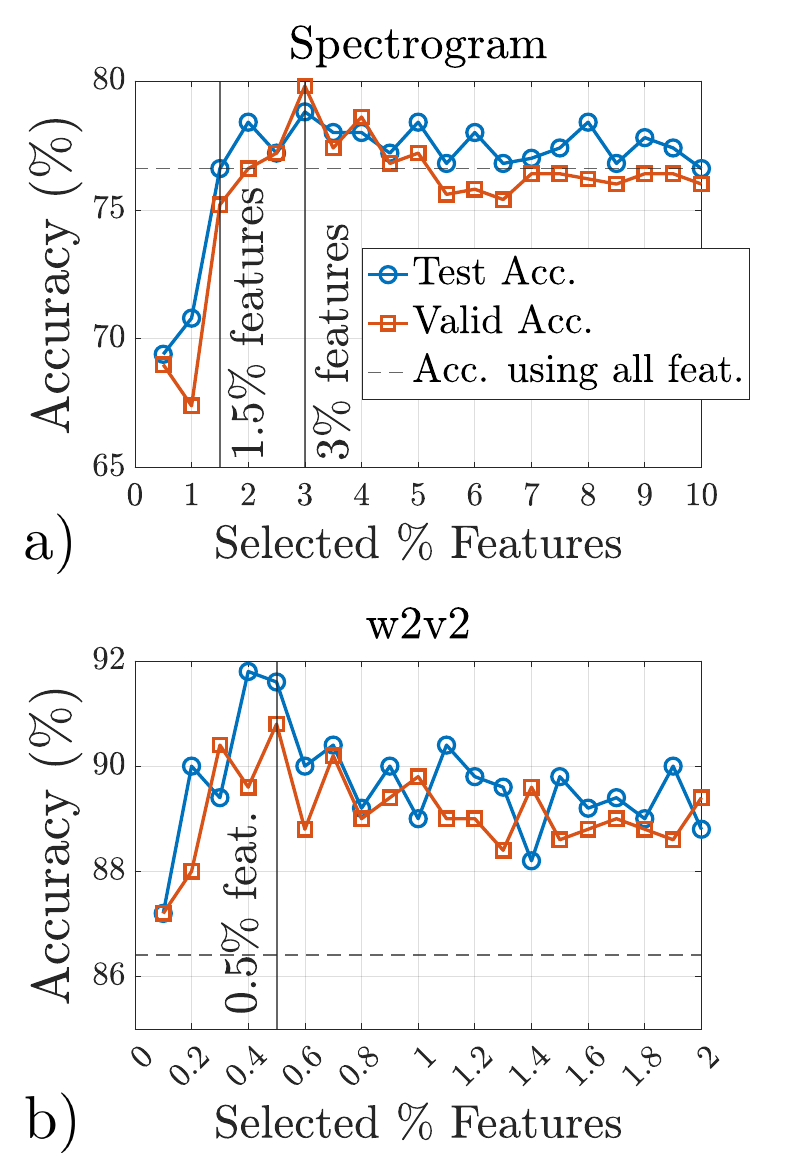}
    \caption{LGBM performance on the validation and test sets using different \% of top-ranked features for (a) spectrogram and (b) w2v2 embeddings. For ease of comparison, the performance on the test set using all features is also illustrated.}
    \vspace{-0.5cm}
    \label{fig:feat_import}
\end{figure}
This difference in the optimal $M$ value for the two representations can be attributed to their different temporal resolution, with a spectrogram frame representing no contextual information and a w2v2 embedding representing contextual information.
As a result, it is expected that correlation should be maximized on a larger time scale for the spectrogram representation than for the w2v2 representation.
In the remainder of this section, we report only results using these optimal chunk sizes computed on the validation set (i.e., \(M=8\) for spectrogram and \(M=24\) for w2v2 representations).

\subsection{MCCA vs PCA}
% \vspace{-0.2cm}
\label{sec:ccapca}

In this section we validate the applicability of MCCA in comparison to the traditional PCA dimensionality reduction for pathological speech detection.
The performance when using MLP and LGBM with different number of PCA or MCCA components on spectrogram and w2v2 representations are presented in Tables~\ref{tab:res_spec} and~\ref{tab:res_w2v2}.

Table~\ref{tab:res_spec} shows that for spectrogram input representations, using MCCA considerably outperforms using PCA, independently of the number of components or classifier used.
These results validate our hypothesis that certain pathology-irrelevant information in spectrogram input representations is temporally uncorrelated, with MCCA suppressing it, and hence, improving the performance of pathological speech detection.
 
It is worth mentioning that using spectrogram input representations with more complex architectures like CNN yields a speaker-level accuracy of $69.72\%$ on the PC-GITA database with the same experimental settings as ours~\cite{janbakhshi_stft}.
Hence, these results also show that using MCCA and simple classifiers such as MLP or LGBM on spectogram input representations results in a considerable performance improvement (of $6\%$ to $7\%$) in comparison to using more complex architectures like CNN.

Table~\ref{tab:res_w2v2} shows that for w2v2 representations, using MCCA yields a similar or a slightly better performance than using PCA, independently of the number of components or classifier used. 

These results reinforce our hypothesis that employing MCCA helps in selecting more relevant
information across time, even when using powerful multilingual SSL embeddings which already yield an impressive performance.

\begin{figure}[ht!]
\vspace*{-0.36cm}
 \hspace*{-0cm}
% \hspace*{cm}  
    \centering
    \includegraphics[scale=0.44]
    {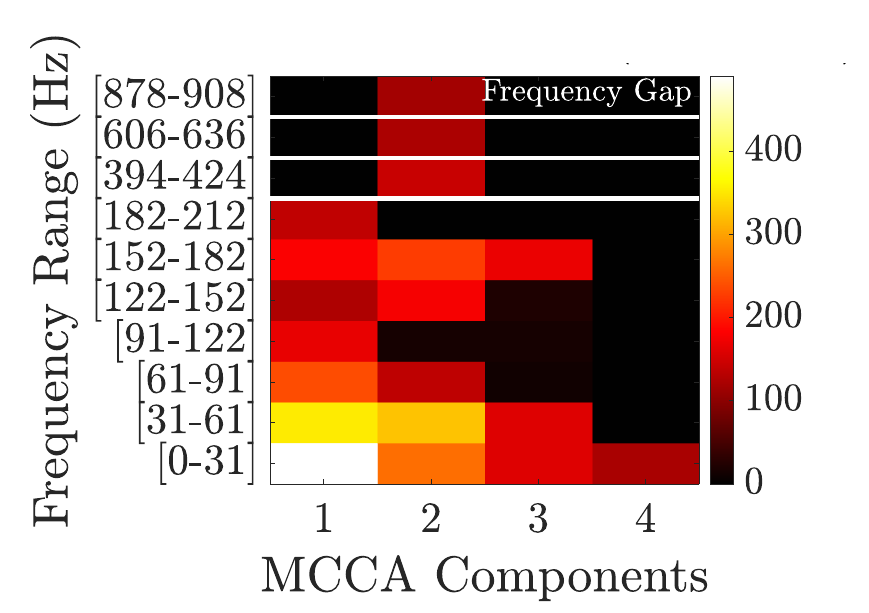}
    \caption{ MCCA components of the frequency bins belonging to the $1.5$\% top-ranked features. The color map illustrates the importance assigned to each bin when using LGBM.} 
    % (a) is zoomed version of (b) just to better visualize the importance of the low freq components.
      \vspace{-0.5cm}
    
    \label{fig:feat_import2}
\end{figure}
\subsection{Important Features and Interpretation}

\label{sec:impo}
Since MCCA is beneficial for pathological speech detection as shown in Section~\ref{sec:ccapca}, in this section we further improve its performance and explore its interpretability. For the analysis presented in this section, we focus on using LGBM and MCCA (with $5$ components) on spectrogram and w2v2 representations.

Using each of the $50$ models ($10$ folds $\times$ $5$ seeds) trained in our previous experiments, we calculate the average importance assigned by LGBM to each of the features across the different training sets. From this ranking, we train new models using only a percentage $k$ of the top-ranked features.
For spectrogram representations, we consider $ k \in [0.5,1,\dots,10]\%$, while for w2v2 representations, we consider $k \in [0.1,0.2,\dots,2]\%$. 
The optimal hyperparameters found in the previous experiment are reused to avoid repeating the grid search. 

Fig.~\ref{fig:feat_import} depicts the performance (on the validation and test sets) obtained when using only the $k$ top-ranked features as well as the test set performance when using all features (previously reported).  

For spectrogram input representations, Fig.~\ref{fig:feat_import}a shows that with only 1.5\% of the features (i.e., $18$ features vs. $32382$ in the spectrogram and $1285$ after MCCA), %(i.e., $18$ features compared to $32,382$ features in the spectrogram and $1285$ after CCA)
we achieve test performance levels similar to using all features.
These $18$ top features are shown in Fig.~\ref{fig:feat_import2}, which depicts the MCCA components only for the frequency bins belonging to the $1.5$\% top-ranked features. It can be observed that these features are located in the bottom-left corner of the MCCA representation, with frequencies ranging from $0$ to $210$ Hz, corresponding to the average frequency of human voice, being highly important as expected.

This plot shows that the choice of MCCA is relevant, as the most correlated features, i.e., component $1$ and $2$, are found to be the most important. 

For w2v2 input representations, Fig.~\ref{fig:feat_import}b shows that using any of the $k$ top-ranked features improves the test performance in comparison to using all features.
More importantly, Fig.~\ref{fig:feat_import} shows that selecting the percentage of features that yields the best validation performance, i.e., $3$\% ($56$) for spectrogram and $0.5$\% ($256$) for w2v2 embeddings, results in an improved test set accuracy, reaching $78.8\%$ and $91.6\%$ respectively. 
These values surpass those obtained in our previous experiments, and notably, exceed what has been reported with such approaches in the literature to the best of our knowledge.
These findings underscore MCCA's effectiveness in extracting meaningful features for PD detection using simple classifiers while preserving the essential feature structure, which is vital for further analysis. Future work will explore the use of the more powerful non-linear MCCA methods proposed in~\cite{kaloga2021variational}.

\section{Conclusion}

In this paper we have proposed to incorporate MCCA in state-of-the-art pathological speech detection approaches based on spectrogram and w2v2 input representations.
The presented results show that MCCA improves the performance through preserving pathology-discriminant cues and discarding pathology-irrelevant information that is uncorrelated across time.
More powerful MCCA methods and generalization and robustness in noisy conditions remain topics to investigate in the future.

%\section*{ACKNOWLEDGEMENTS}

  % \newpage
\bibliographystyle{IEEEtran}
\bibliography{mybib}

\end{document}